\documentclass[fleqn,usenatbib]{rasti}

\usepackage{newtxtext,newtxmath}

\usepackage[T1]{fontenc}

\DeclareRobustCommand{\VAN}[3]{#2}
\let\VANthebibliography\thebibliography
\def\thebibliography{\DeclareRobustCommand{\VAN}[3]{##3}\VANthebibliography}

\usepackage{graphicx}	
\usepackage{amsmath}	

\newcommand{\kk}{\text{K\textsubscript{gal}}}

\title[Short title, max. 45 characters]{A quantum-enhanced support vector machine for galaxy classification}

\author[M. Hassanshahi et. al.]{
Mohammad Hassan Hassanshahi,$^{1}$\thanks{E-mail: m.hassanshahi@ucl.ac.uk}
Marcin Jastrzebski,$^{1}$
Sarah Malik$^{1}$
and Ofer Lahav$^{1,2}$
\\
$^{1}$Department of Physics and Astronomy, University College London, Gower St, London, WC1E 6BT, UK\\
$^{2}$Center for Data Intensive Science and Industry, University College London, Gower St, London, WC1E 6BT, UK\\
}

\date{Accepted XXX. Received YYY; in original form ZZZ}

\pubyear{2022}

\begin{document}
\label{firstpage}
\pagerange{\pageref{firstpage}--\pageref{lastpage}}
\maketitle

\begin{abstract}

Galaxy morphology, a key tracer of the evolution of a galaxy's physical structure, has motivated extensive research on machine learning techniques for efficient and accurate galaxy classification. The emergence of quantum computers has generated optimism about the potential for significantly improving the accuracy of such classifications by leveraging the large dimensionality of quantum Hilbert space. This paper presents a quantum-enhanced support vector machine algorithm for classifying galaxies based on their morphology. The algorithm requires the computation of a kernel matrix, a task that is performed on a simulated quantum computer using a quantum circuit conjectured to be intractable on classical computers. The result shows similar performance between classical and quantum-enhanced support vector machine algorithms. For a training size of $40$k, the receiver operating characteristic curve for differentiating ellipticals and spirals has an under-curve area (ROC AUC) of $0.946\pm 0.005$ for both classical and quantum-enhanced algorithms. Additionally, we demonstrate for a small dataset that the performance of a noise-mitigated quantum SVM algorithm on a quantum device is in agreement with simulation. Finally, a necessary condition for achieving a potential quantum advantage is presented. This investigation is among the very first applications of quantum machine learning in astronomy and highlights their potential for further application in this field.
\end{abstract}




\section{Introduction}
Studying the morphology of galaxies is essential for understanding their evolution and formation. Therefore, classifying galaxies based on their morphology is crucial in observational cosmology. Such classification was initially performed by Hubble~\citep{hubble1926extragalactic}, in which bulge-dominant and disk-dominant galaxies were differentiated. With the increase of galaxy images collected by sky surveys, automated algorithms, such as machine learning (ML), were developed to achieve high-speed and accurate galaxy classification. One of the first examples of using ML to classify galaxies was presented by~\citet{lahav1995galaxies} while more recent examples include \citet{CompanySVM2008} which uses support vector machine algorithm and \citet{cheng2020optimizing} which compares several ML techniques.

The emergence of quantum computers and their remarkable progress in recent years has raised hopes that these machines can potentially boost the performance of ML techniques, thanks to their distinct characteristics compared to classical computers. In this new computing paradigm, classical bits are replaced with quantum bits (qubits), which can interfere and entangle each other. A classical computer must keep track of $2^n$ parameters to execute a quantum computing algorithm with $n$ well-entangled (non-separable) qubits. On the other hand, such an algorithm is executed naturally and hence more efficiently on a quantum computer. Therefore, an ML algorithm enhanced by quantum computing could potentially outperform fully-classical ML algorithms. With the help of quantum computers, such algorithms can be executed efficiently, while not all of these algorithms are guaranteed to be efficiently executable by classical computers alone.

Support vector machine (SVM)~\citep{SVM_book} is a supervised classification algorithm that finds a hyperplane separating data into two classes. The remarkable feature of SVM is the use of the \textit{kernel method}, which facilitates non-linear classification by implicitly mapping data to a (typically higher-dimensional) space where a linear classification can be performed. With the kernel method, and where the noise-level is sufficiently low, data could be mapped to the exponentially large Hilbert space of qubits using quantum circuits in a way which is intractable to classical computers~\citep{havlivcek2019supervised}. Such quantum circuits are made of a set of unitary quantum gates which can rotate single or multiple qubits by an amount controlled by the value of data points. An advantage of the kernel method is that an explicit mapping of all data points to the feature space is not needed. Instead, it is sufficient to compute their inner product and construct the corresponding \textit{kernel matrix}. This implicit mapping can significantly reduce computation costs.

A necessary condition for a possible quantum advantage is to use a quantum circuit that is difficult to simulate on classical computers. Although there is a long list of such circuits, they typically include many layers of quantum gates (a.k.a~\textit{deep} circuits) such that they are not implementable on currently-available quantum computers. These quantum computers, named \textit{near-term} or \textit{noisy intermediate-scale quantum (NISQ)} devices, are prone to noise which can significantly deteriorate the performance of deep quantum circuits. Recently,~\citet{havlivcek2019supervised} proposed a family of quantum circuits with an intermediate depth which is not only implementable on near-term devices but is also conjectured to be difficult to simulate on classical computers. 

In this study, we developed an SVM algorithm to classify elliptical and spiral galaxies based on their shapes. The kernel matrix was computed in two different ways: (i) using (simulated) quantum computers and (ii) using classical computers. The classifiers produced with these kernels are called quantum and classical kernel classifiers hereafter. The quantum kernel was computed using a quantum circuit from the quantum circuit family~\citep{havlivcek2019supervised} which was conjectured to achieve a quantum advantage. The kernel was then fed into the classical SVM optimiser for finding the classifier hyperplane. In contrast, the conventional SVM algorithm was performed for the classical kernels. Given the random nature of quantum processes, the quantum kernel can only be \textit{estimated} by measuring the qubits R times on a real quantum computer. Since the wave function amplitudes are accessible in simulation, we computed the exact kernel matrix (or equivalently estimated it for $R{=}\infty$ in a noiseless quantum computer).

As mentioned earlier, noise is a major problem in near-term devices. One of the main advantages of quantum-enhanced SVM algorithms is that when the quantum circuit used for mapping the features in the kernel method is not too deep, error mitigation techniques can be applied~\citep{temme2017error, li2017efficient, kandala2019error, liu2021rigorous}. Such techniques allow the algorithm to be executed on near-term devices without significant loss of computational power.  The effectiveness of error-mitigation techniques and the robustness of kernel entries to noise have been demonstrated on near-term devices~\citep{havlivcek2019supervised, kusumoto2021experimental, bartkiewicz2020experimental, supernova-qke, liu2021rigorous}. These advantages make quantum-enhanced SVM algorithms leading candidates for achieving quantum advantage on near-term devices~\citep{liu2021rigorous}.

Quantum machine learning is new in the field of astronomy. \citet{caldeira2019restricted} have used Restricted Boltzmann Machines (RBMs) for a morphology classification of galaxies using a quantum annealer. They found that for small datasets, a quantum annealer-based RBM outperforms certain classical algorithms. In another study, \citet{supernova-qke} have employed the SVM algorithm with quantum kernel estimation to classify two supernova types. They designed a quantum circuit which is robust for execution on near-term devices although the quantum circuit used in the study is not difficult to simulate on classical computers. The paper demonstrated that the classification performance on a near-term device is comparable to the noiseless simulation. 

Section~\ref{sec:dataproc} illustrates the procedure for the collection and pre-processing of the input data, section~\ref{sec:svm} describes the SVM algorithm and quantum kernel estimation, and section~\ref{sec:result} explains the results of this study. 

\section{The galaxy data}
\label{sec:dataproc}
\begin{figure}
	\centering	
    \includegraphics[width=1.5in]{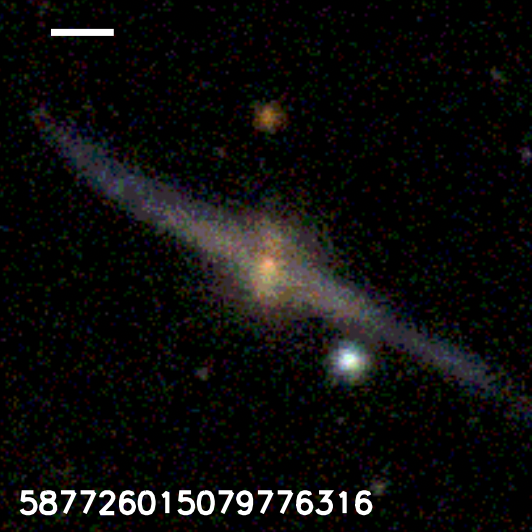}
    \includegraphics[width=1.5in]{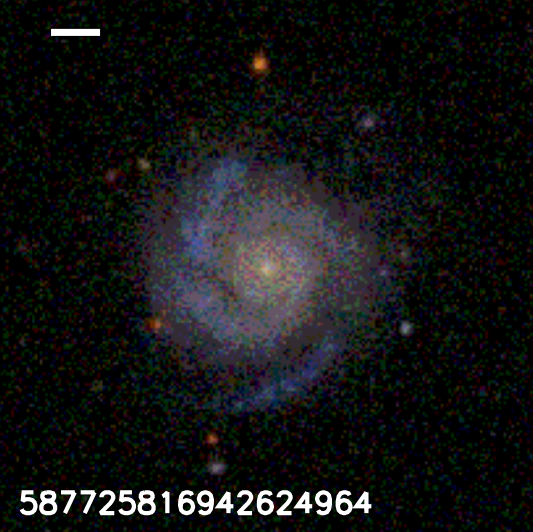}
    \includegraphics[width=1.5in]{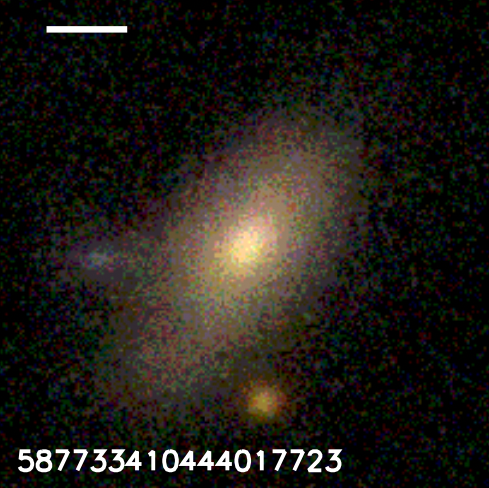}
    \includegraphics[width=1.5in]{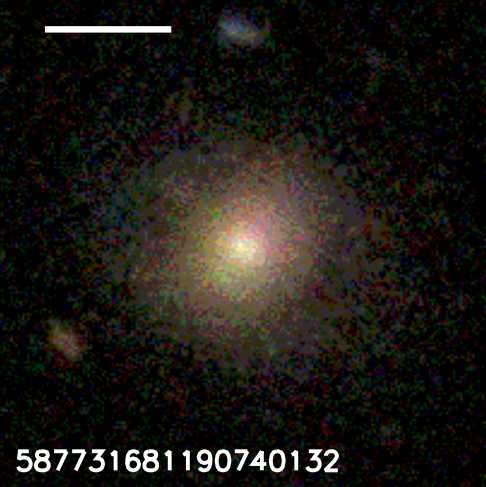}
    
    \caption{Examples of spiral (top row) and elliptical (bottom row) galaxies. The images were taken from~\citet{schawinski2010galaxy}}
    \label{fig:glximage}
\end{figure}

The data used in this analysis is collected from the publicly available Galaxy Zoo 1 (GZ1) dataset~\citep{galaxyZoo1, galaxyZoo}. This dataset includes morphological classification of galaxy images drawn from the Sloan Digital Sky Survey (SDSS). A few examples of these images, which are 2D projections of 3D bodies, are shown in Fig.~\ref{fig:glximage}. A large number of volunteers contributed to this classification by labelling galaxies visually based on their shapes. After performing bias corrections, galaxies are classified as `spiral' or `elliptical' if more than 80\% of the debiased votes are in these categories, while all other galaxies are labeled as `uncertain'. These labels are considered true labels throughout this study.

The features of the galaxies in this dataset are collected from the morphological metrics provided in a catalog in~\citet{comparativeML}. A total of five distance-independent features are included in this analysis for model training:  
(i) concentration $C = \log_{10}(R_{out}/R_{in})$~\citep{conselice2003relationship, lotz2004new, ShannonHferrari2015}, where $R_{out}$ and $R_{in}$ are the radii of the spheres enclosing (in this measurement) 75\% and 25\% of the total galaxy flux, 
(ii) asymmetry $A=1-s(I^0-I^\pi)$ defined using the Spearman's rank correlation coefficient $s()$ of the flux of the galaxy image $I^0$ and its $\pi$-rotated version $I^\pi$, 
(iii) smoothness $S=1-s(I^0-I^S)$ defined similarly for a comparison with the flux of the smoothed image, 
(iv) second gradient moment G2 extracted from the Gradient Pattern Analysis (GPA) method~\citep{rosa2018gpa}, and 
(v) the Shannon information entropy H of the galaxy image pixels~\citep{ShannonHferrari2015} which is expected to be low for smooth galaxies. More information about these features can be found in~\citet{comparativeML}. All features were extracted using CyMorph~\citep{rosa2018gpa, comparativeML}, written in Cython, a language for writing C extensions for Python.

The features dataset was merged with the GZ1 dataset by matching the ID of galaxies. The merged dataset was trimmed by removing galaxies for which non-physical values were assigned or the CyMorph algorithm failed. Galaxies labelled `uncertain' were also removed from the dataset, leaving only `spiral' and `elliptical' labels. Features were normalised by linearly scaling them to the [0,1] interval.

A property defined in~\citet{comparativeML} for each galaxy image is the area of the galaxy derived from its Petrosian radius $R_p$ divided by the area of the point spread function (PSF) estimated from the full width at half maximum (FWHM), 
\begin{equation}
\kk = (\frac{R_p}{\text{FWHM}/2})^2 .
\end{equation}
(More information on $R_p$ can be found in \citet{petrosian1976surface, petrRadEisen2011}.) Samples with large \kk{} generally include larger objects. The dataset used in this analysis has a spatial resolution of $0.396$ arcsec/pixel and a PSF FWHM of ${\approx}$1.5 arcsec~\citep{comparativeML}. This mediocre resolution led us to divide the dataset into $\kk \ge 5$, $\kk \ge 10$, and $\kk \ge 20$ and train a classifier for each, following what is done in \citet{comparativeML}.
Since \kk{} has a similar distribution for the ellipticals and spirals, and to avoid unnecessary divergence from~\citet{comparativeML}, it was not used in the classifiers as an input feature. 

\section{Machine learning model}
\label{sec:svm}
\subsection{Support Vector Machine}
\begin{figure}
    \includegraphics[width=3.4in]{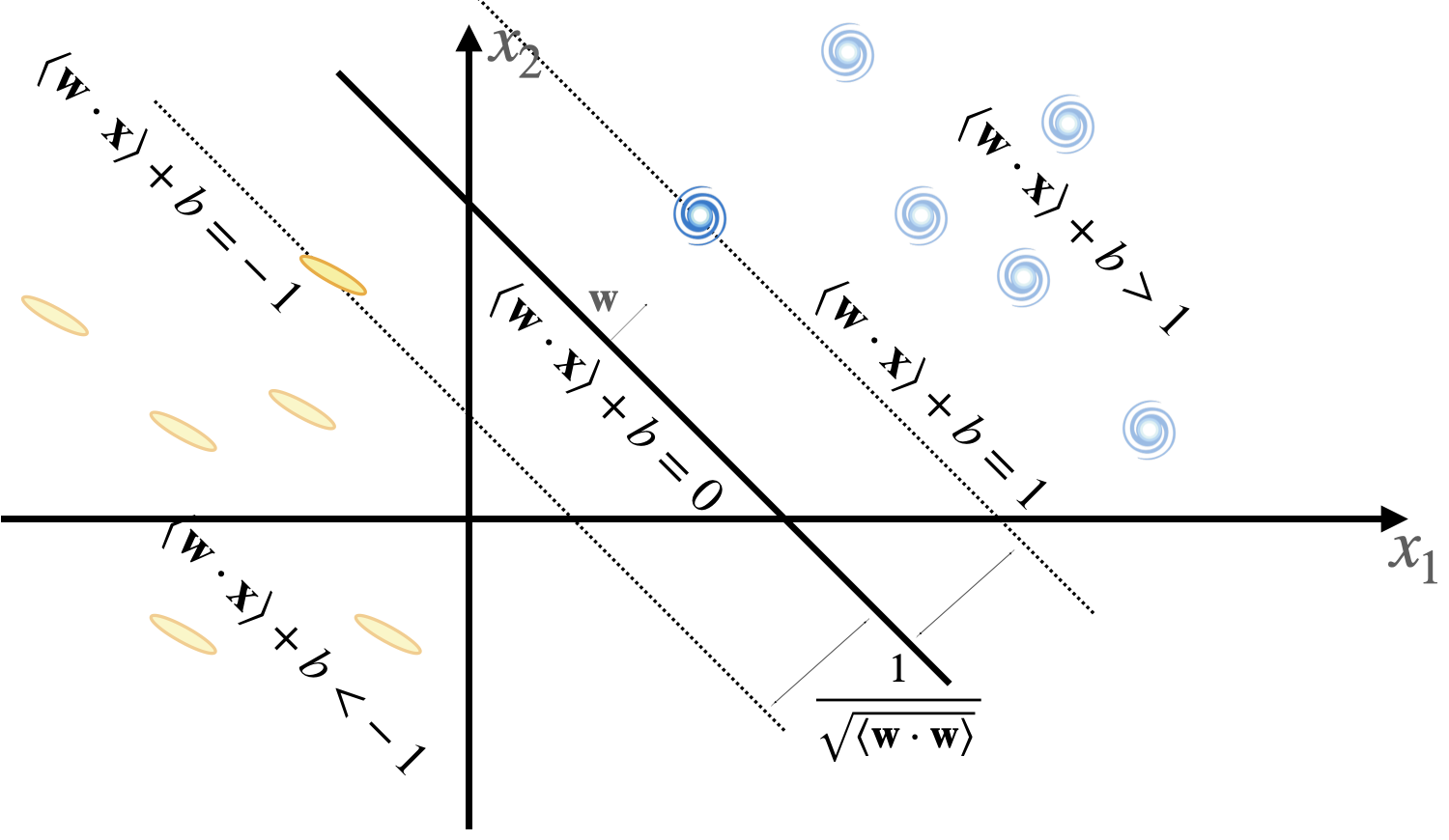}
    \caption{Illustration of how the SVM algorithm separates the elliptical (yellow) and spiral (blue) galaxies in the space spanned by two features $x_1$ and $x_2$. The decision boundary hyperplane is $\mathbf{w}^\intercal \mathbf{x} + b = 0$, and the closest data points to the hyperplane, which are located on $\mathbf{w}^\intercal \mathbf{x} + b = \pm 1$, are called support vectors.}
    \label{fig:svm}
\end{figure}

In order to classify galaxies, we utilized the support vector machine (SVM) algorithm. For a dataset ($\mathbf{x}_1$, $y_1$), ..., ($\mathbf{x}_M$, $y_M$), this algorithm finds a hyperplane which maximises the separation (margin) between two classes, where $\mathbf{x}_i{\in}\mathbb{R}^N$ and $y_i{\in}\{-1,1\}$ are the feature vector and the corresponding class label for the i'th datapoint, respectively. The hyperplane can be formulated as $\mathbf{w}^\intercal \mathbf{x} + b = 0$, where $\mathbf{w}$ is the normal vector of the hyperplane and $b/\lVert \mathbf{w} \rVert$ is the distance of the origin to the hyperplane. The nearest datapoints $\mathbf{x}_i$ to the hyperplane from either of classes are called support vectors (SV) and the margin boundaries are hyperplanes passing through SVs, $\mathbf{w}^\intercal \mathbf{x} + b = \pm 1$ (see Fig.~\ref{fig:svm}). When the two classes are linearly separable, all datapoints $\mathbf{x}_i$ with $y_i{=}1$ ($y_i{=}-1$) satisfy $\mathbf{w}^\intercal \mathbf{x} + b \ge 1$ ($\mathbf{w}^\intercal \mathbf{x} + b \le -1$), meaning that they lie on the correct side of the margin (a.k.a~hard margin). However, a more relaxed margin condition (a.k.a.~soft margin) can be used in SVM, which allows some data points of each class to cross their corresponding margin boundary in exchange for a penalty term in the loss function. Taking this into account, the SVM optimisation problem can be formulated as
\begin{equation}
\begin{aligned}
\min_ {w, b, \zeta} \frac{1}{2} \lVert \mathbf{w} \rVert ^2  + C \sum_{i=1}^{M} \zeta_i\\
\textrm{subject to: } & y_i (\mathbf{w}^\intercal \mathbf{x}_i + b) \geq 1 - \zeta_i, \\
							  & \zeta_i \geq 0, \textrm{ and}\\
							  & \forall i \in \{1, ..., M\},
\end{aligned}
\end{equation}
where $\zeta_i$ is the distance of $\mathbf{x}_i$ to its corresponding margin boundary if it has crossed the boundary and $\zeta_i{=}0$ otherwise. $C$ is a hyperparameter of the optimization problem, which controls the strength of the penalty for the crossed-boundary data points.

This optimisation problem is the primal representation of a dual problem which can be shown to be
\begin{equation}
\begin{aligned}
\min_{\alpha} \frac{1}{2} \boldsymbol{\alpha}^\intercal \mathbf{Q} \boldsymbol{\alpha} - \mathbf{e}^\intercal \boldsymbol{\alpha}\\
\textrm{subject to: } & \mathbf{y}^\intercal \boldsymbol{\alpha} = 0,\\
							  & 0 \leq \alpha_i \leq C, \textrm{ and}\\
							  & \forall i \in \{1, ..., M\},
\end{aligned}
\end{equation}
where $\mathbf{e}$ is a vector of ones and $\mathbf{Q}$ is an $M\times M$ matrix with $\mathbf{Q}_{ij} \coloneqq y_i y_j K(\mathbf{x}_i, \mathbf{x}_j)$, where the \textit{kernel matrix} $K$ is constructed by the inner product of datapoints, $K(\mathbf{x}_i, \mathbf{x}_j) = \langle \mathbf{x}_i, \mathbf{x}_j\rangle $. After the optimisation is complete, the class of a new datapoint $\mathbf{x}_k$ is predicted using the sign of the decision function 
\begin{equation}
\label{eq:decisionFunction}
y_k = \mathrm{sgn}(\sum_{i\in SV} y_i \alpha_i K(\mathbf{x}_i, \mathbf{x}_k) + b).
\end{equation}

\begin{figure}
    \includegraphics[width=3.1in]{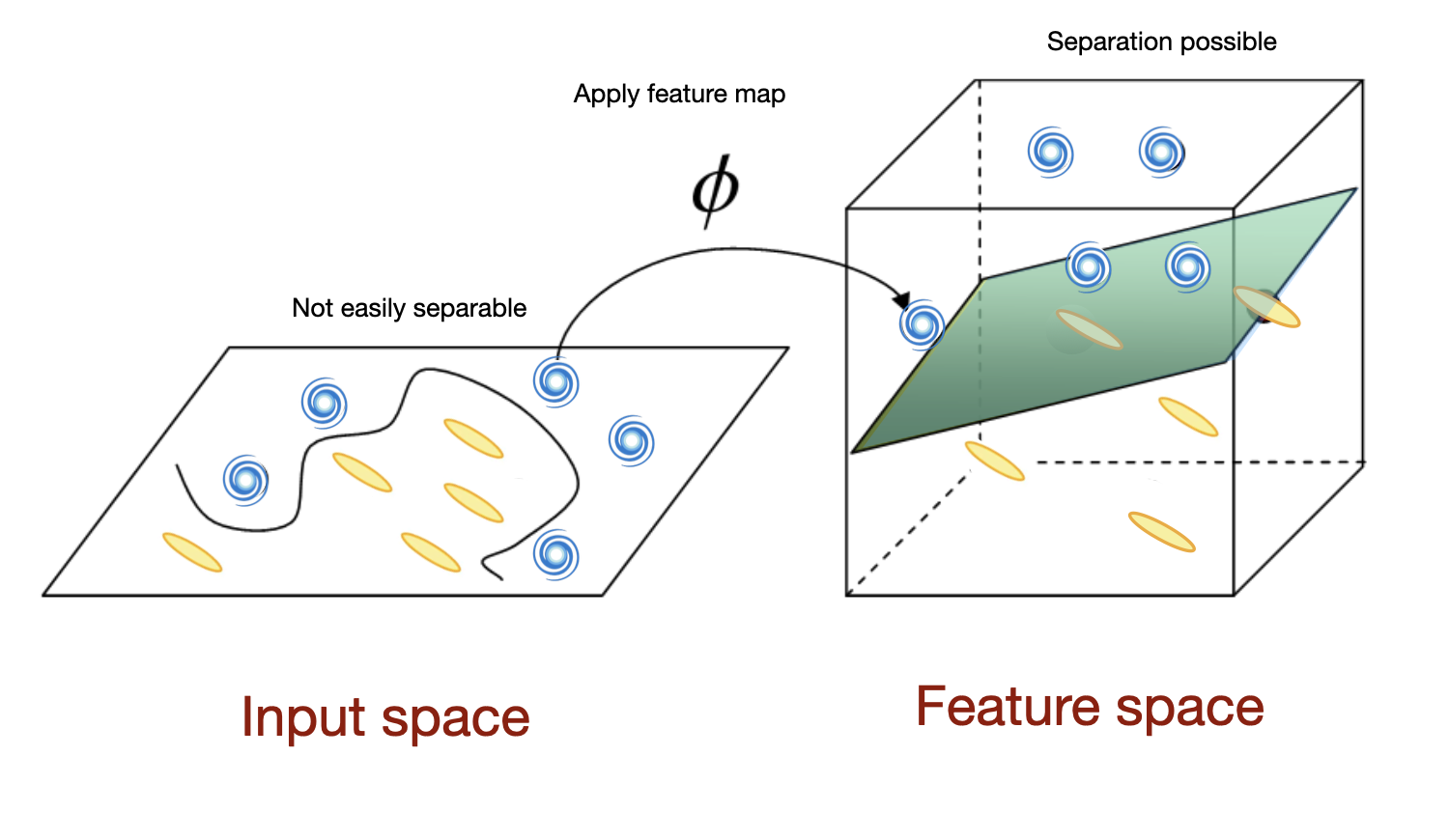}
    \caption{Illustration of kernel method. When data are not linearly separable in the input space, they are mapped to the feature space and linearly classified.}
    \label{fig:featureMap}
\end{figure}

The kernel matrix can be utilized for an efficient non-linear data classification with the \textit{kernel method} . In this method, before computing the inner products, datapoints are mapped with a \textit{feature map} into a high-dimensional space, called \textit{feature space}, where a linear classification is performed (see Fig.~\ref{fig:featureMap}). The kernel matrix therefore becomes $K(\mathbf{x}_i, \mathbf{x}_j) = \langle \phi(\mathbf{x}_i), \phi(\mathbf{x}_j)\rangle $.

An essential property of the kernel method is that it only requires computing inner products, thereby avoiding an explicit mapping of the datapoints to the feature space that may be computationally expensive. 

\subsection{Quantum Kernel Estimation}

Quantum computers can be used to estimate the kernel matrix in the SVM algorithm. It is based on the idea that instead of conventional classical feature spaces, one can exploit the exponentially large Hilbert space by leveraging controllable entanglement and superposition between qubits. A datapoint $\mathbf{x}_i$ is non-linearly mapped to the quantum state $\rho (\mathbf{x}_i)=|\psi(\mathbf{x}_i)\rangle \langle \psi(\mathbf{x}_i)|$ in the Hilbert space, which serves as our feature space. In this space, an inner product between two quantum states $\rho (\mathbf{x}_i)$ and $\rho (\mathbf{x}_j)$ is defined by tracing over their product, and each kernel matrix entry is subsequently calculated to be
\begin{equation}
K(\mathbf{x}_i, \mathbf{x}_j) = \mathrm{tr}\{\rho (\mathbf{x}_i)\rho (\mathbf{x}_j)\} = |\langle \psi(\mathbf{x}_i)|\psi(\mathbf{x}_j)\rangle | ^2.
\end{equation}
This matrix entry can be calculated using unitary matrix $\mathcal{U}$:
\begin{equation}
\label{eq:uu}
|\langle \psi(\mathbf{x}_i)|\psi(\mathbf{x}_j)\rangle | ^2 = |\langle 0^{\otimes N}|\mathcal{U^\dagger}(\mathbf{x}_i) \mathcal{U}(\mathbf{x}_j) |0^{\otimes N}\rangle |^2,
\end{equation}
where $|0^{\otimes N}\rangle$ is the initial state with all qubits in the $|0\rangle$ state. The kernel matrix is then estimated by measuring the state $\mathcal{U^\dagger}(\mathbf{x}_i) \mathcal{U}(\mathbf{x}_j) |0^{\otimes N}\rangle$ R times (shots) and then calculating the fraction of times where all qubits are measured to be 0. In our study, following the design of the circuit family proposed in~\citet{havlivcek2019supervised}, the number of qubits N was chosen to be the same as the number of features. We computed the asymptotic case (R=${\infty}$) by accessing the output of the simulated quantum circuit.

Depending on the choice of feature maps, and hence $\mathcal{U}$, the complexity of estimating the quantum kernel and the performance of the resulting classifier changes. To achieve quantum advantage, we are particularly interested in feature maps based on quantum circuits that cannot be simulated efficiently on classical computers, while maintaining high performance for the classifier. The number of parameters a classical computer needs to track grows exponentially with the number of qubits if the qubits are well entangled. This can make the simulation of quantum circuits difficult, especially when the circuit is sufficiently \textit{deep} including a large number of intermediate gates, and this is where the quantum advantage is achievable.  Current quantum computers are not large enough to implement a deep quantum circuit for this purpose. However, a recent study~\citep{havlivcek2019supervised} proposed a quantum circuit conjectured to provide quantum advantage while, more importantly, being implementable on current quantum computers. The authors showed that estimating kernel with the quantum circuit described below is directly related to a 3-fold forrelation (`fourier correlation')~\citep{forrelation} problem and could lead to quantum advantage.

A more generalised version of the unitary $\mathcal{U}$ proposed in~\citet{havlivcek2019supervised} is available as \textit{PauliFeatureMap} class in the Qiskit software development kit.  We used a subset of the generalised unitary which are of the form $\mathcal{U}(\mathbf{x}) = U_{\phi(\mathbf{x})} H^{\otimes N} U_{\phi(\mathbf{x})} H^{\otimes N}$, where $H$ is the conventional Hadamard gate which puts computational bases into equal superposition while $U_{\phi(\mathbf{x})}$ is an entangling unitary parametrised by datapoint $\mathbf{x}$.

\begin{figure}
    \includegraphics[width=3.1in]{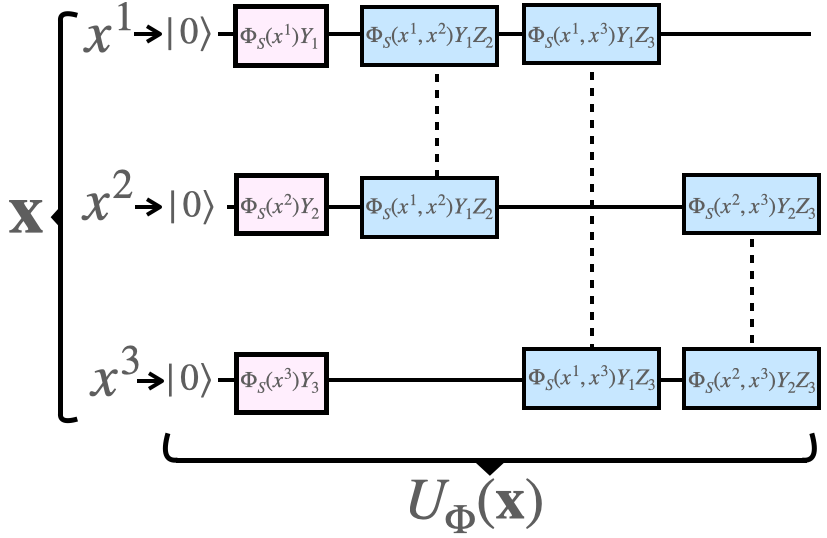}
    \caption{Quantum circuit for an example of unitary $U_{\phi(\mathbf{x})}$ with three qubits, each taking one of the features $x^i$ of datapoint $\mathbf{x}$ as input. Single-qubit and two-qubit gates are shown in pink and blue, respectively.}
    \label{fig:unitary_example}
\end{figure}

\begin{figure}
    \includegraphics[width=3.1in]{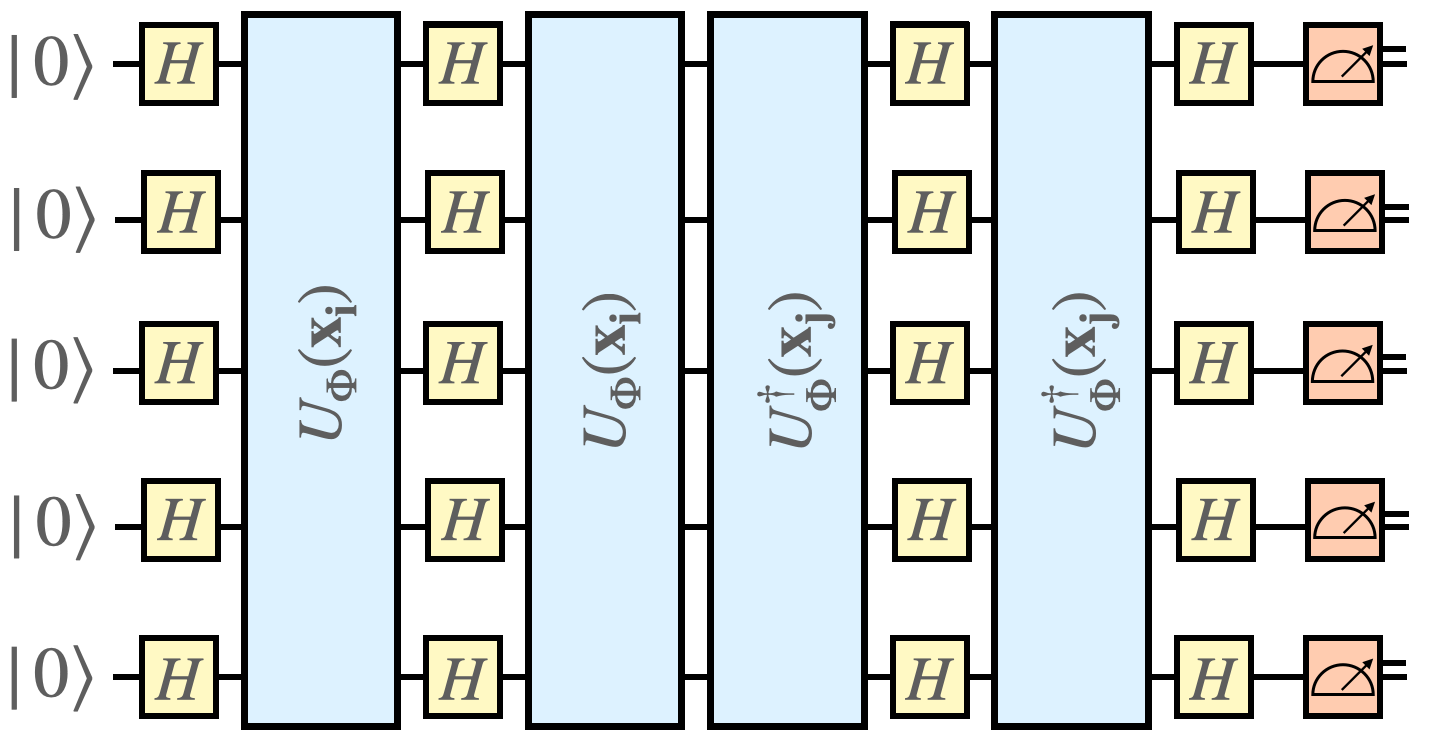}
    \caption{The full quantum circuit used for estimating the inner product $|\langle \psi(\mathbf{x}_i)|\psi(\mathbf{x}_j)\rangle | ^2$. Unitaries $U_{\phi(\mathbf{x})}$ (blue) are interleaved with Hadamard gates (yellow) and the quantum state is measured (orange) at the end of the circuit.}
    \label{fig:circuit}
\end{figure}

The unitary $U_{\phi(\mathbf{x})}$ is formulated as
\begin{equation}
\label{Eq:u_phi}
U_{\phi(\mathbf{x})}=\exp\left(i\;\alpha\sum_{S\subseteq [N]}
\phi_S(\mathbf{x})\prod_{i\in S} P_i\right),
\end{equation}
where $\alpha$ controls rotations and interactions, $P_i \in \{ I, X, Y, Z \}$ denotes the identity and Pauli matrices, $S {\in} \{\binom{n}{k}\ \text{combinations},\ k {=}1,... n \}$ shows a subset of qubits (or features) to interact, and $\phi_S(\mathbf{x})$ is a user-defined function of features which adjusts the amount of rotation. In this study, we only considered $|S| {\le} 2$, meaning that three- and more qubit interactions are excluded. Fig.~\ref{fig:unitary_example} presents an example of this unitary for the case where single-qubit rotation $P_i {=} Y$ and two-qubit interaction $P_{j,k} {=} YZ$ while no ${\ge} 3$-qubit interactions exist. The full circuit used for our kernel estimation is shown in Fig.~\ref{fig:circuit}. 

After the kernel is estimated, it is fed into a conventional SVM to find the optimised hyperplane. Once the hyperplane is found, datapoints from the test set are classified using the decision function in Eq.~\ref{eq:decisionFunction}.

\section{Classification results}
\label{sec:result}
This section describes the result of using our classical and quantum kernel classifiers to predict the morphological types of galaxies. An important step for minimising the loss function of a classifier is hyperparameter optimisation. For each kernel, the hyperparameters which maximise the area under the receiver operating characteristic curve (ROC AUC) for separating ellipticals and spirals were found using a grid search.

In the classical kernel, the commonly used radial basis function (RBF) kernel was used. The regularisation term $C$ was searched over 9 orders of magnitude, $10^s$ for $-1 \le s\le 8$, while for the kernel coefficient $\gamma$, the search included values between $0.0001$ and $100$, with most values centred around $1$. During the search, hyperparameter configurations that significantly underperformed compared to other configurations were removed. A full grid search was then performed on the remaining parameters. The best hyperparameter for the classical kernel was $\gamma=0.01$ and $C=10^7$.

A similar approach was taken for the quantum kernel. The hyperparameters in this kernel are different unitaries $U_{\phi(\mathbf{x})}$ defined in Eq.~\ref{Eq:u_phi}. Circuits with single-qubit rotations followed by two-qubit interactions between all pairs of qubits are considered in our study, where the rotations and interactions are induced by Pauli matrices (an example shown in Fig.~\ref{fig:unitary_example}). The rotation factor $\alpha$ changed from $0.005$ to $1.4$ while the regularisation term $C$ varied between $10$ and $10^8$. We used the default data-mapping function $\phi_S(\mathbf{x})$ of the PauliFeatureMap class, which is $\phi_S(\mathbf{x})=x_0$ for single-qubit rotations (i.e. when $|S|=1$) and $\phi_S(\mathbf{x})=(\pi-x_0)(\pi-x_1)$ for two-qubit interactions (i.e. when $|S|=2$). The configuration which yielded the best ROC AUC score was $\alpha=0.03$, $C=10^7$, and unitary $U_{\phi(\mathbf{x})}$ with $P_i {=} Y$ and $P_{j,k} {=} YZ$.

\begin{figure}
    \includegraphics[width=3.1in]{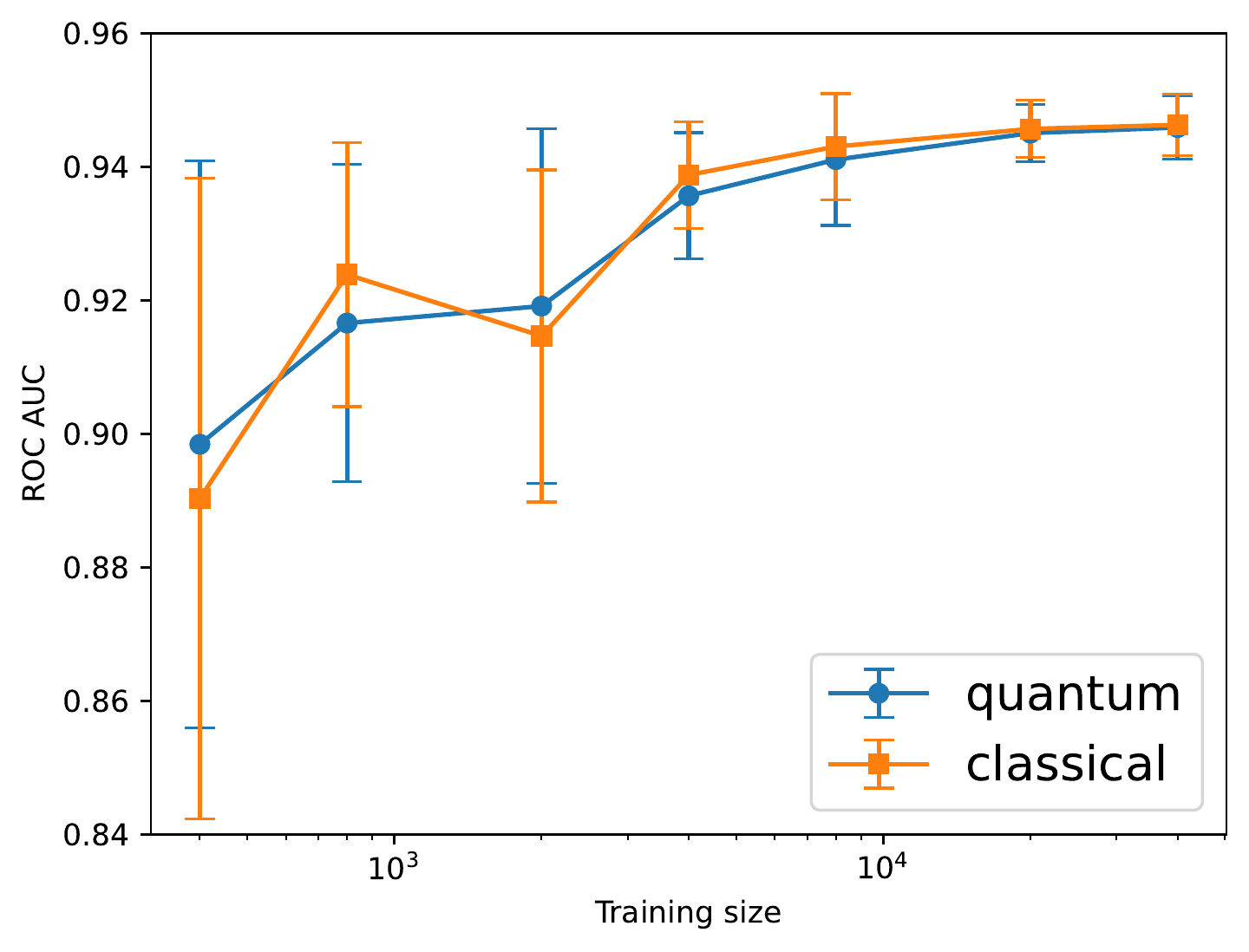}
    \caption{ROC AUC score as a function of training size for quantum and classical kernel classifiers. The ROC scores are computed by applying the models to the test sets using 5-fold cross-validation. The data for this plot is for $\kk \ge 5$ condition.}
    \label{fig:rocVsTrainSize}
\end{figure}

The ROC AUC score of the quantum and classical kernel classifiers are compared as a function of training size in Fig.~\ref{fig:rocVsTrainSize}. The scores derived for the $\kk \ge 5$ dataset show that the two classifiers have a comparable performance regardless of the training size. We utilised 40k data points to train the classifiers, a number constrained by the computational resources available for running the quantum kernel classifier. Additionally, 10k datapoints were allocated for testing. A ROC AUC score of $0.946\pm 0.005$ was obtained with both the classical and quantum kernels. The two classifiers could also achieve a high ROC AUC score for small training sizes. As expected, the ROC AUC score and its uncertainty are improved with the training size, almost reaching a plateau at the training size of 40k. 

Since the dataset used in this study is derived from~\citet{comparativeML}, it is worth comparing the results. The authors in that study used a boosted decision tree (BDT) and a deep learning (DL) method for performing spiral-elliptical binary classification of galaxies. While this comparison provides a general understanding of the relative performance of the classifiers, it might not be a fair comparison due to subtle differences. For example, (i) the training size in our study was limited by computing resources required for the quantum kernel classifier, (ii) we applied 5-fold cross-validation while they used an 80-10-10 splitting of the data, and (iii) we optimised the hyperparameters by maximising the ROC AUC score while they potentially chose another score. The comparison is summarised in Table~\ref{tab:bdtVsDL}. When galaxies with low $\kk{}$ values are included in the dataset ($\kk \ge 5$), the SVM outperforms the BDT, while its performance is worse than the DL. For large values of $\kk{}$ ($\kk \ge 20$), the performance of the BDT and SVM are comparable, but both are worse than the DL. Given the above-mentioned differences between the two studies, the main point of this comparison is that the results are compatible with each other.

\begin{table}
 \caption{ROC score for different models and conditions.}
 \label{tab:bdtVsDL}
 \begin{tabular}{lccc}
  \hline
  							  									& $K\ge 5$  				& $K\ge 10$ 				& $K\ge 20$  \\
  \hline
  BDT~\citep{comparativeML}						&0.901 						& 0.928 					& 0.976			\\
  DL~\citep{comparativeML}						& 0.971 					& 0.977 					& 0.985			\\
  Classical SVM (This study) 						& 0.946 					& 0.962 					& 0.975			\\
  Quantum-enhanced SVM (This study) 		& 0.946 					& 0.961						& 0.975			\\
  \hline
 \end{tabular}
\end{table}

\begin{center}
\begin{figure}
	\centering
    \includegraphics[width=2.7in]{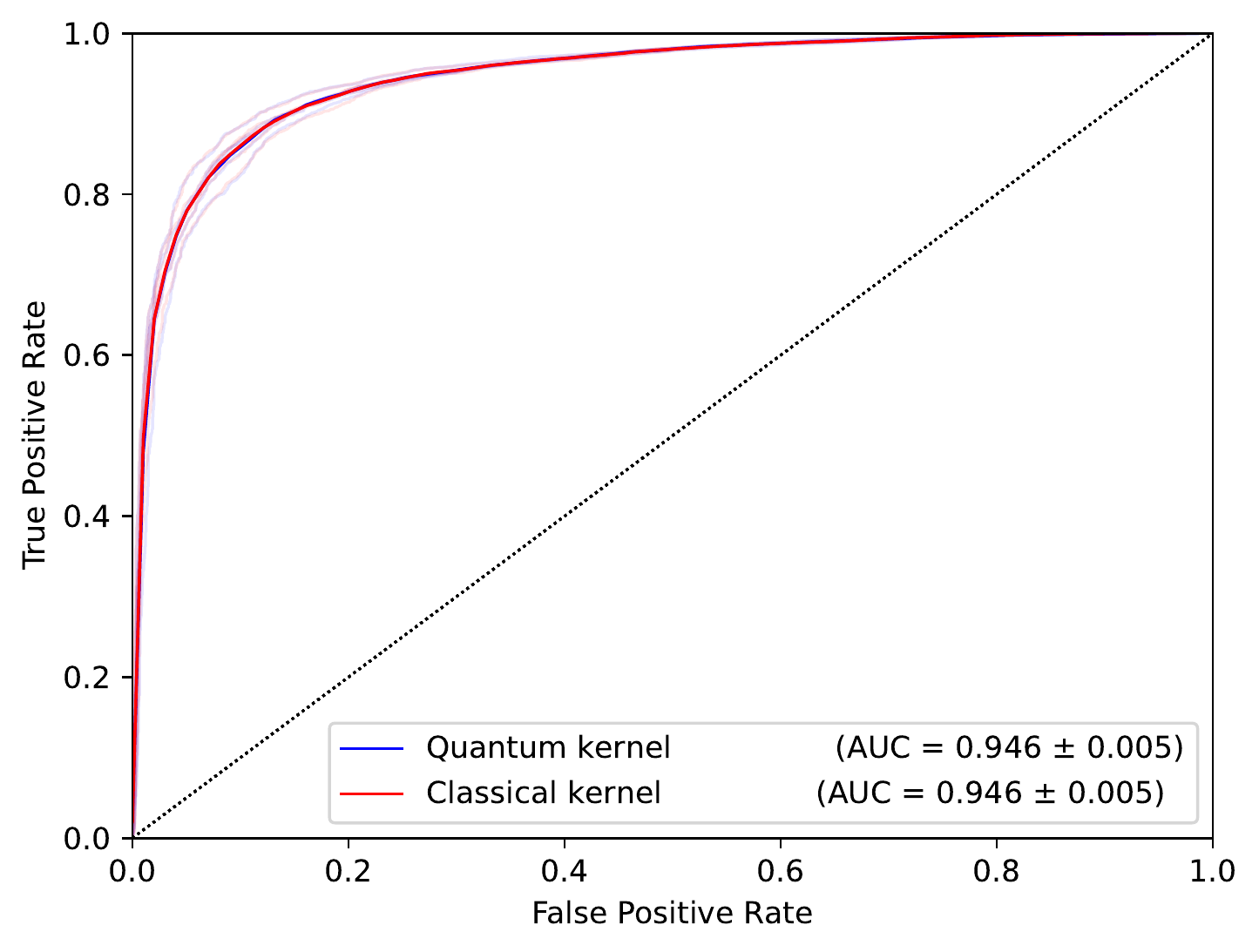}
    \par\bigskip
    \includegraphics[width=2.7in]{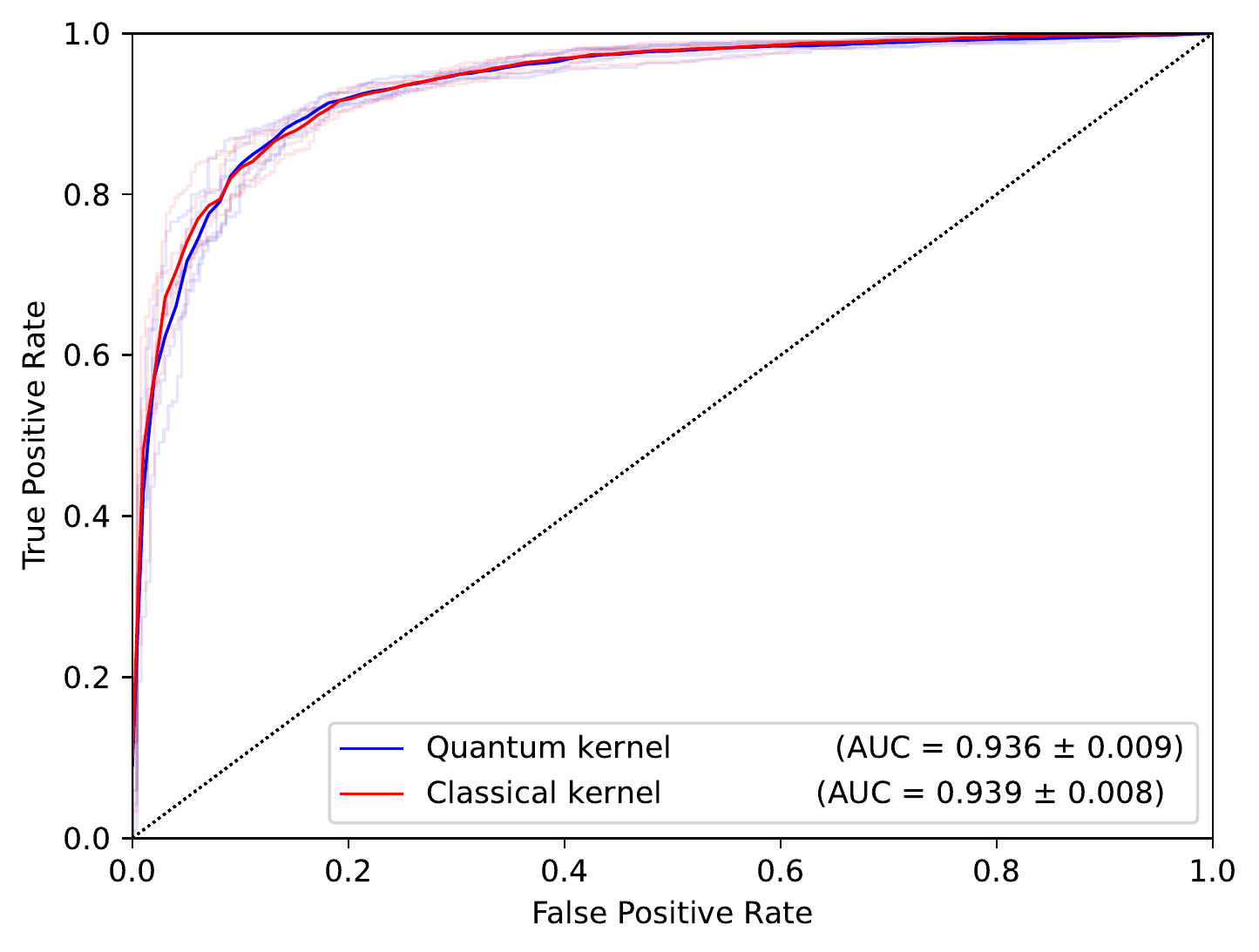}
    \par\bigskip
    \includegraphics[width=2.7in]{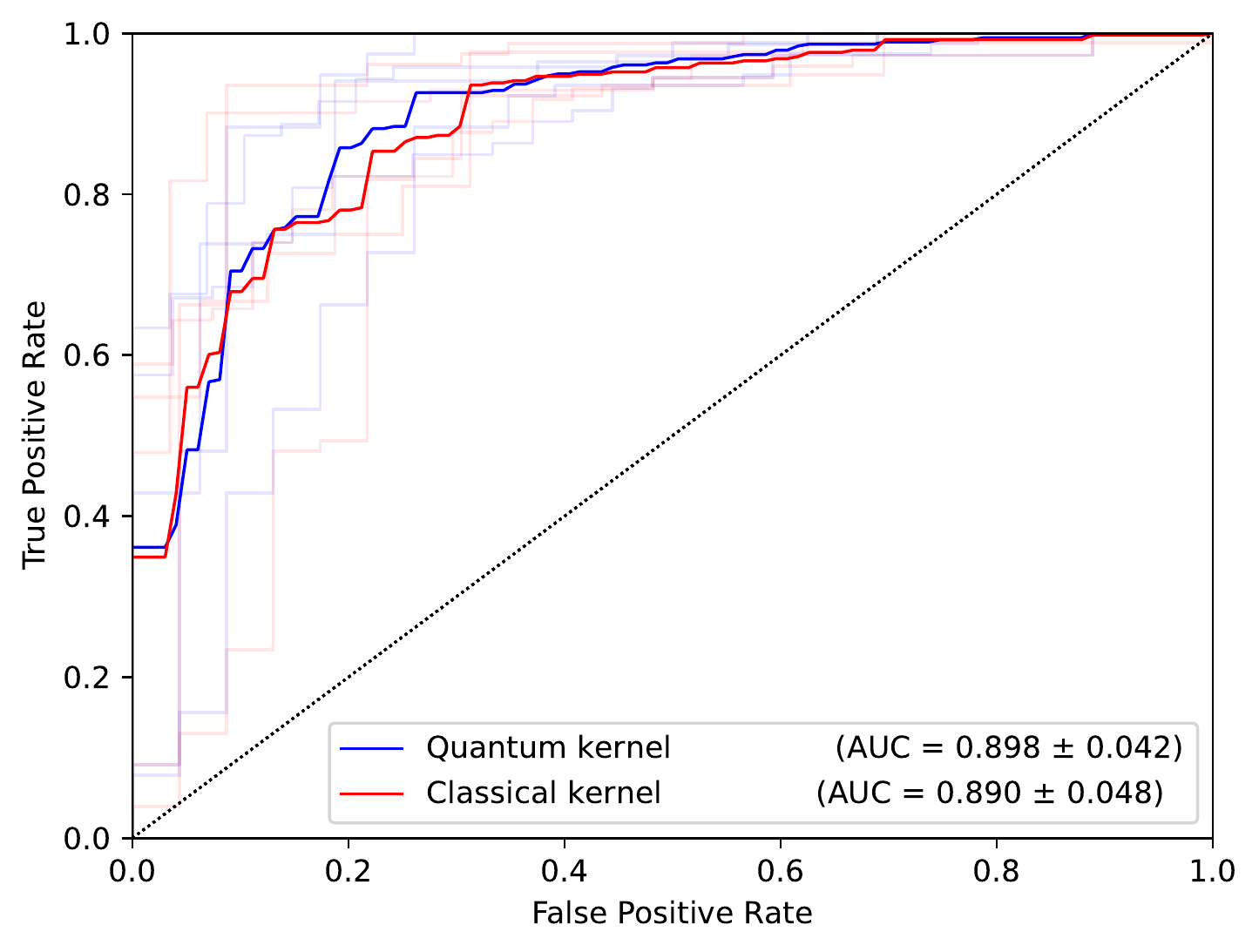}
    \caption{ROC curves along with the ROC AUC scores of the quantum (blue) and classical (red) kernel classifiers for the training sizes of 40000 (top), 4000 (middle), and 400 (bottom). The uncertainty of the ROC AUC scores is the standard deviation of the corresponding 5-fold ROC AUC scores. The ROC curves per fold are plotted using light colors. These plots correspond to the dataset with $K \ge 5$.}
    \label{fig:rocCurve_K5}
\end{figure}
\end{center}

The ROC curves for the quantum and classical kernel classifiers are compared in Fig.~\ref{fig:rocCurve_K5}. The curves are for the $\kk \ge 5$ dataset for different training sizes. The two kernels exhibit comparable performance.

\begin{figure}
    \includegraphics[width=3.1in]{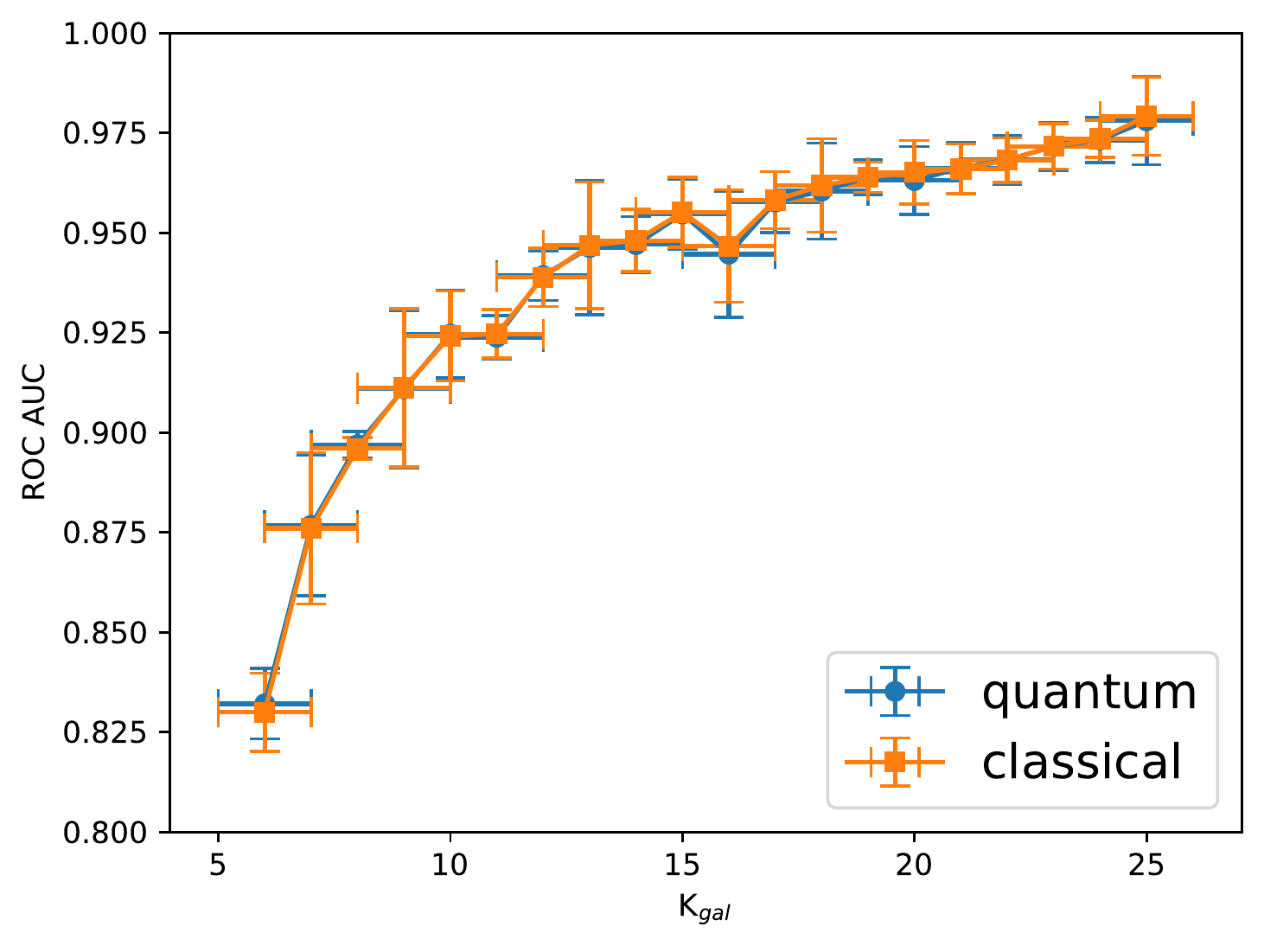}
    \caption{ROC AUC as a function of $\kk{}$. These ROC AUC scores and their uncertainties are derived from performing a 5-fold cross-validation on 50000 galaxy images. The model trained on $5 \le \kk{}$, $10 \le \kk{}$, and $20 \le \kk{}$ is applied to the $5 \le \kk{} \le 10$, $10 \le \kk{} \le 20$, and $20 \le \kk{}$ regions, respectively.}
    \label{fig:rocAUCvsK}
\end{figure}

As mentioned in Sec.~\ref{sec:dataproc}, the value of $\kk{}$ is proportional to the size of galaxies. Comparing the performance of the classifiers as a function $\kk{}$ may indicate if one kernel type can extract more information from the shape of the galaxies for certain galaxy size ranges. Fig.~\ref{fig:rocAUCvsK} displays the ROC AUC score as a function of $\kk{}$ for the classifiers developed with the quantum and classical kernels. The performance of the two classifiers is similar in all ranges of $\kk{}$.

\subsection*{Running on hardware}
\begin{figure}
    \includegraphics[width=3.1in]{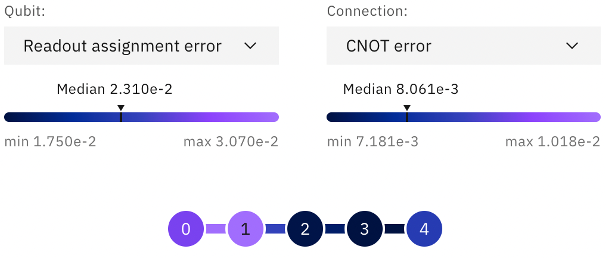}
    \caption{A schematic of the \textit{ibmq\_manila} quantum device. The qubits are shown with numbered circles, with colors indicating their readout assignment error. Where a line connects two qubits, a (multi-qubit) gate can be simultaneously applied on both. The line colors show the error produced after applying a controlled-NOT (CNOT) gate on the qubit pairs.}
    \label{fig:manila}
\end{figure}

In order to demonstrate the feasibility of executing the algorithm on a quantum device, we implemented the algorithm on the IBM \textit{ibmq\_manila} device (see Fig.~\ref{fig:manila}). This device consists of 5 qubits which are positioned in a line, each qubit connected to its nearby qubit(s). Multi-qubit interactions are allowed only between connected qubits. For instance, a two-qubit gate can be applied to the qubit pair 2 \& 3 as they are connected, whereas for the qubit pair 2 \& 4, the gate must be preceded by a \textit{Swap} gate on the pair 2 \& 3 or 3 \& 4. While the Swap gate facilitates a connection between any two qubits as long as there is a connection path between them, the gate adds a relatively large amount of noise to the device and hence should be avoided in near-term devices. In order to avoid using Swap gates, we simplified the kernel circuit such that two-qubit gates are applied only to nearby qubits, as opposed to every-pair of qubits described previously. To further reduce the noise level, we used dynamical decoupling~\citep{dynamicalDecoup} technique for \textit{error suppression} and twirled readout error extinction (T-REx) technique~\citep{TRex-errorMitig} for \textit{error mitigation}. More information about error-handling in quantum devices can be found on the Qiskit website~\citep{QiskitError}.

Constructing a kernel matrix for training an SVM model requires calculating $N_\text{train}\times (N_\text{train}-1)/2$ matrix entries (i.e.~between each pair of training datapoints). Additionally, when applying a trained model to a test set, $N_\text{train}\times N_\text{test}$ matrix entries must be calculated (i.e.~between each training and testing datapoints). For calculating each matrix entry, the corresponding circuit needs to be executed a large number of times (shots) to reduce sampling uncertainty. Our choice of $N_\text{train}$, $N_\text{test}$, and the number of shots has been optimised to fit the limitations of the quantum device. We selected 50 spirals and 50 ellipticals at random before merging them into a dataset of size 100. Since the original dataset is imbalanced, an equal selection of spirals and ellipticals makes the ROC curves smoother and allows us to better compare the performance of simulation and quantum device. A 4-fold cross validation is performed on the dataset with $N_\text{train}=75$ and $N_\text{test}=25$ while the number of shots is set to 4000. 

Table~\ref{tab:device} shows the ROC AUC score for the quantum device and simulation. In simulation, $N_\text{train}$, $N_\text{test}$, and the number of shots were set equal to the ones used for the quantum device. The performance of the simulation and quantum device agree within the uncertainties. As the dataset size is small, the uncertainties of the scores are relatively large. Therefore, the similarity between simulation and quantum device is not necessarily extendible to much larger datasets.

\begin{table}
 \caption{The ROC AUC score of the quantum device and simulation for a dataset of size $N_\text{train}=75$ and $N_\text{test}=25$. For each circuit corresponding to a kernel entry, the number of shots was set to 4000.}
 \label{tab:device}
 \begin{tabular}{lccc}
  \hline
  							  			& Mean 					& Std.~Dev. 		\\
  \hline
  Quantum device				&0.83 					& 0.10 			\\
  Simulation						&0.80 					& 0.08 			 \\
  \hline
 \end{tabular}
\end{table}

\section{Potential quantum advantage}
The limited size of current quantum computers prevents an experimental comparison between quantum devices and conventional computers on the runtime of quantum algorithms. However, the time complexity of the algorithms can be estimated with theory, as has been done in~\citet{duckett2022reconstructing}. 

In the quantum SVM algorithm, an $N\times N$ kernel matrix is constructed in the training process, where $N$ is the number of datapoints in the training set. This requires $\mathcal{O}(\beta N^2)$ calculations, where $\beta$ is a coefficient which can depend on different factors such as the number of features $M$, the solution accuracy $\epsilon$, as well as $N$.

When the quantum kernel is computed on a quantum device, the value of $\beta$, symbolized as $\beta_Q$, has been shown~\citep{havlivcek2019supervised, gentinetta2022complexity} to be proportional to $\epsilon^{-2}$. It has also been shown~\citep{gentinetta2022complexity} that $\beta_Q$ scales with the training size as $N^{8/3}$ due to sampling uncertainty. We can therefore conclude that $\beta_Q = \mathcal{O}(N^{8/3}\epsilon^{-2})$. Given a fixed accuracy $\epsilon$, a potential quantum advantage is achieved if for certain number of features and datapoints $\beta_Q<\beta_C$, where $\beta_C$ is the value of $\beta$ for a classical simulation of the quantum kernel. The current best classical algorithm proposed in~\citet{bravyi2021classical} leads to $\beta_C = \mathcal{O}(2^{2M/3}\epsilon^{-2/3})$, which when compared to $\beta_Q$ has a better scaling with $N$ and $\epsilon$ but an exponentially worse scaling with $M$. This means that a quantum advantage would be possible for large number of features as described below.

\begin{figure}
    \includegraphics[width=3.1in]{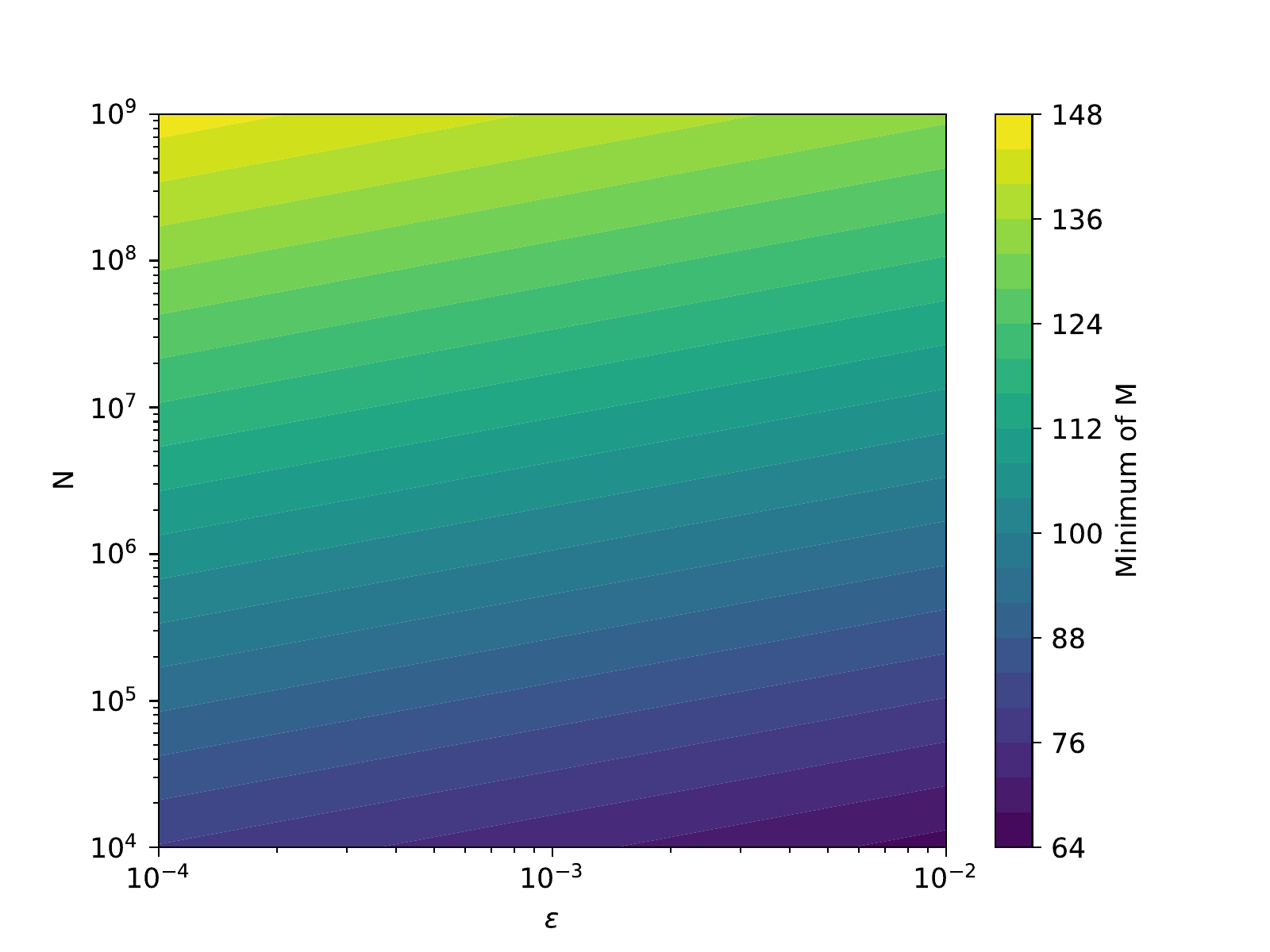}
    \caption{An approximated minimum number of features (M) needed for the quantum kernel algorithm to perform faster on a quantum device compared to the simulation.}
    \label{fig:M-N-E}
\end{figure}

Fig.~\ref{fig:M-N-E} shows an approximate minimum number of features required to achieve a potential quantum advantage, as a function of $\epsilon$ and $N$. The plot is derived by equating $\beta_Q=\beta_C$ and the approximation comes from the fact that orders of magnitude (rather than absolute values) were used for $\beta$. Requiring an accuracy of $0.001$, the minimum number of features varies from $73$ to $140$ when the training size increases from $10^4$ to $10^9$ galaxies. 

Whether or not a quantum advantage in kernel computation is \textit{practically} achievable is yet an open question~\citep{schuld2022quantum, exponentialConcentration}. For instance, it is possible that further studies introduce additional overhead for $\beta_Q$, rejecting a quantum advantage for this algorithm. Our study serves as a proof-of-concept, demonstrating an initial promise of using quantum kernel for galaxy classification, while also underlining the necessity for future comprehensive examinations of potential limitations and challenges. 

\section{Conclusions}
In this study, we apply a quantum-enhanced SVM algorithm for classifying galaxies into spirals or ellipticals based on their morphology. The Galaxy Zoo 1 dataset was used to collect volunteer-labelled galaxies, and five features per galaxy were extracted from the catalogue provided in~\citet{comparativeML}. The SVM algorithm utilises the kernel method by implicitly mapping data points into a feature space (with a feature map), where they are classified with a hyperplane. In this algorithm, quantum computers can be used to estimate the kernel matrix, which is then passed to a standard SVM optimiser to find the optimal hyperplane using classical computers. For this galaxy classification problem, we employed a feature map that is feasible for implemention on near-term quantum computers and is conjectured to be intractable on classical ones. Using simulations, we showed that the performance of this algorithm is comparable to a fully-classical SVM for training sizes ranging from $400$ to $40$k. Following~\citet{comparativeML}, we used a parameter \kk{}, which is proportional to the size of galaxies, to compare the performance of our classical and quantum SVM algorithms for different ranges of this parameter. Both algorithms exhibit similar improvement in performance with increasing \kk{}. For the dataset with $\kk \ge 5$ and the training size of $40$k, the ROC AUC score was found to be $0.946\pm 0.005$ for both the classical and quantum kernel classifiers, where the uncertainty is the standard deviation of the scores derived from 5-fold cross-validation. We also executed a slightly simplified version of the algorithm on an IBM quantum device and showed that the result is compatible with simulation within uncertainties.

Our findings show that, despite the limited number of qubits provided by current devices, quantum models can provide similar performance to classical ones across a wide range of training sizes. This is in agreement with previous studies~\citep{supernova-qke, belis2021higgs, wu2021application, fadol2022application, duckett2022reconstructing, schuhmacher2023unravelling}. It has recently been shown~\citep{park2020practical} that a quantum SVM is capable of achieving higher performance for datasets with complex boundaries between the two classes. Future studies could explore whether incorporating a larger number (or different set) of features provides a different class boundary, leading to an improvement in the performance of the quantum classifier compared to the classical one. Given that our quantum circuit requires an equal number of qubits and features, adding features requires additional qubits. However, the rapid proliferation of available quantum devices makes the issue of qubit availability less concerning for future work. Another possibility to improve the performance of the quantum classifier is to encode the relationships (if present) between galaxy features in the quantum circuit (e.g. see \citet{heredge2021quantum}). Future studies could also investigate how to perform multi-class classification with a reasonable training dataset.

Future theory and experimental studies should shed light on the possibility of achieving a practical quantum advantage with the SVM algorithm. Further investigation of different noise contributions could indicate a worse time complexity as well as a reduced performance for the quantum algorithm. While acknowledging these potential challenges, we showed that based on the currently-available scaling estimation, the minimum number of galaxy features necessary for a potential quantum advantage scales logarithmically with the dataset size. With the availability of $\mathcal{O}(100)$-qubit near-term devices in the near future~\citep{ibmChallenge100}, we encourage future research to investigate the performance of quantum SVM algorithms with larger number of features extracted from galaxy images. It should be emphasised that the quantum kernel used in our study can accommodate any number of features and datapoints.

In conclusion, our result motivates further study of quantum machine learning techniques to problems in astronomy.

\section*{Acknowledgements}

OL acknowledges STFC consolidated and CDT-DIS grants for their support of this work. SM and MH are funded by grants from the Royal Society. We also acknowledge funding from the STFC.

\section*{Data Availability}

The Galaxy Zoo 1 (GZ1) dataset was derived from the official Galaxy Zoo website \url{https://data.galaxyzoo.org/} with the direct link \url{https://galaxy-zoo-1.s3.amazonaws.com/GalaxyZoo1_DR_table2.csv.gz}. The features were taken from the supplementary material of~\citet{comparativeML}, available directly in \url{https://www.sciencedirect.com/science/article/pii/S2213133719300757#ec-research-data}.



\bibliographystyle{rasti}
\bibliography{example} 




%
%


\bsp	
\label{lastpage}
\end{document}